\documentclass[preprint,onecolumn,nofootinbib,superscriptaddress,showkeys,citeautoscrip]{revtex4}
\pdfoutput=1
\usepackage[usenames,dvipsnames]{color}
\usepackage{graphicx}
\usepackage{amsmath}
\usepackage{amsfonts}
\usepackage{amssymb}
\usepackage{dsfont}
\usepackage{dcolumn}
\usepackage{bm}
\usepackage{slashed}
\usepackage{pstricks}
\usepackage{slashed}
\usepackage{multirow}
\usepackage{array}
\usepackage{rotating}
\usepackage{xcolor}
\usepackage{epstopdf}
\usepackage{fancybox}
\usepackage{ulem,fancyvrb}

\newcolumntype{x}[1]{
{\centering}p{#1}}%

\newcommand{\GeV}      {~\mathrm{GeV}}

\def \cha{\widetilde{\chi}^{\pm}_1}

\newcommand{\beqn}{\begin{eqnarray}}
\newcommand{\eeqn}{\end{eqnarray}}
\newcommand{\be}{\begin{equation}}
\newcommand{\ee}{\end{equation}}
\newcommand{\non}{\nonumber \\}

\newcommand{\mathsym}[1]{{}}

\def \cha{\tilde{\chi}^{\pm}_1}

\def \na{\tilde{\chi}^{0}_1}

\def \n34{\tilde{\chi}^{0}_{3,4}}
\def \g{\tilde{g}}

\def \ta{\tilde{t}_1}

\def \ba{\tilde{b}_1}

\def \sta{\tilde{\tau}_1}

\def \sml{\tilde{\mu}_L}
\def \smr{\tilde{\mu}_R}
\def \ser{\tilde{e}_R}
\def \sel{\tilde{e}_L}

\def\met100{\slashed{E}_T\geq 100 \GeV}

\newcommand{\gappeq}{\mathrel{\rlap {\raise.5ex\hbox{$>$}}
{\lower.5ex\hbox{$\sim$}}}}
\newcommand{\lappeq}{\mathrel{\rlap{\raise.5ex\hbox{$<$}}
{\lower.5ex\hbox{$\sim$}}}}

\def\met{\slashed{E}_{T}}
\def\meff{m_{\rm eff}}

\begin{document}

\title{New Constraints on Dark Matter from CMS and ATLAS data}

\author{Sujeet~Akula}
\affiliation{Department of Physics, Northeastern University,
 Boston, MA 02115, USA}

\author{Daniel~Feldman}
\affiliation{Michigan Center for Theoretical Physics,
University of Michigan, Ann Arbor, MI 48109, USA}

\author{Zuowei~Liu}
\affiliation{C.N.\ Yang Institute for Theoretical Physics, 
Stony Brook University, Stony Brook, NY 11794, USA}

\author{Pran~Nath}
\affiliation{Department of Physics, Northeastern University,
 Boston, MA 02115, USA}

\author{Gregory~Peim}
\affiliation{Department of Physics, Northeastern University,
 Boston, MA 02115, USA}


\begin{abstract}
Constraints on dark matter  from the first CMS and ATLAS SUSY searches 
 are  investigated.  It is shown that within the minimal supergravity model, 
the early search for supersymmetry at the LHC has depleted a large portion of 
the signature space in dark matter direct detection experiments.
In particular, the prospects for detecting signals of dark matter in the XENON 
and CDMS experiments are significantly affected in the low 
neutralino mass region.  Here the relic density of dark matter
typically arises from slepton coannihilations in the early universe.
In contrast, it is found that the CMS and ATLAS analyses leave untouched
the Higgs pole and the  Hyperbolic Branch/Focus Point
 regions, which are now being probed by the most recent XENON results.
Analysis is also done for supergravity models with non-universal
soft breaking   
where one finds that a part of the dark matter signature space
depleted by the CMS and ATLAS cuts
in the minimal SUGRA case is repopulated. Thus, observation of
dark matter in the LHC  depleted region of minimal supergravity
may indicate non-universalities in soft breaking.
\end{abstract}

\maketitle


\section{Introduction}
CMS and ATLAS have recently reported their first results for supersymmetry searches~\cite{cmsREACH,AtlasSUSY,atlas0lep}
and have put new constraints on the parameter space of the  $\mathcal{N}=1$ supergravity unified model~\cite{sugra} which,
with  universal boundary conditions  on the  soft breaking parameters at the unification scale, is the model mSUGRA~\cite{sugra,hlw,ArnowittNath}. 
In a subsequent work~\cite{Akula:2011zq},  the implications of the CMS and ATLAS searches on the mSUGRA parameter space was analyzed
in the context of indirect constraints  from LEP and Tevatron searches, from the Brookhaven $g_{\mu}-2$ experiment,
from FCNC constraints in B-physics, i.e., $b\to s\gamma$ and $B_s^0\to\mu^+\mu^-$
and from WMAP.  Some related works have appeared in~\cite{related}.

In this work we analyze the impact of the first results from CMS and ATLAS SUSY  searches on
the direct detection of dark matter~\cite{xenon,cdms}. 
It is found that the LHC results have a large impact on the signature space available
for the  low mass slepton coannihilation region, depleting a significant region where 
direct detection experiments are sensitive to detecting a signal. 
 Thus, we explore the effect of the recent LHC data 
  on the prospects for directly detecting cold dark matter in experiments
  such as XENON and CDMS in supergravity unified models.
  We will discuss both minimal supergravity models, and 
  SUGRA models with 
  non-universal soft breaking terms at the grand unification scale.
  
  For completeness,
  we begin with a brief summary of the independent parameters generated
  by softly broken supergravity theories which are needed to test such models
  at colliders and in dark matter experiments. 
    Comprehensive reviews can be found in~\cite{sugraR,KaneFeldman,Hunt}.
The conditions under which the soft breaking in the minimal 
supergravity model are derived are summarized as follows:  
(i)  supersymmetry is broken through a super Higgs effect giving mass to the gravitino
through the presence of a hidden sector (singlet);
(ii) the hidden and the visible interact only gravitationally;
(iii) the K\"{a}hler potential is generation independent;
(iv) the gauge kinetic function  is minimally  linear in the hidden sector singlet.
This then gives rise to soft terms of the form~\cite{sugra} 
\begin{eqnarray}
{\cal L}_{soft} &=&- \frac{1}{2}(M_a 
\lambda^a \lambda^a + h.c.)
- m_{\alpha}^2 C^{*\alpha}
C^{\alpha}
\nonumber\\ &&
-\
\left(\frac{1}{6} A_{\alpha \beta \gamma} Y_{\alpha \beta \gamma}
	    C^{\alpha} C^{\beta} C^{\gamma}
  + B_0 \mu_0
   H_1 H_2+h.c.\right)\ ,
\end{eqnarray}
where $\lambda^a$ are the gauginos,  $ H_{i=1,2}$ are Higgs doublets, and $C^{\alpha}$
are  the slepton, squark and Higgs fields 
of the minimal supersymmetric standard model.
For the case of universal boundary conditions at the unification (GUT) scale, 
    $m_{\alpha} = m_0$ is the universal scalar mass, $M_a=
    m_{1/2}$ is the universal gaugino mass, $A_{\alpha\beta\gamma} =A_0$  is the universal trilinear coupling, and
    $B_0\mu_0$ is the bilinear coupling where $\mu_0$ is the Higgs mixing parameter that 
    enters the superpotential in the form $\mu_0H_1H_2$
     (all at the GUT scale). 
Thus, the minimal supergravity models are specified by the following set of 
GUT scale parameters ($m_0, m_{1/2}, A_0, B_0,\mu_0$). 
The renormalization group improved scalar potential at the electroweak symmetry breaking scale $Q$ is given by 
 \begin{eqnarray}
\label{loophiggs}
 V&=&m_1^2|H_1|^2 +m_2^2 |H_2|^2 -m_3^2 (H_1H_2+h.c.) \nonumber\\ 
&+& \frac{(g_2^2+g_Y^2)}{8} (|H_1|^2-|H_2|^2)^2 + \Delta V_1,\nonumber\\
 \Delta V_1&=& \frac{1}{64\pi^2}
 \sum_a (-1)^{2s_a}(2s_a+1) M_a^4 \left [ \ln \frac{M_a^2}{Q^2}- \frac{3}{2}\right ] ~,
 \end{eqnarray}
where the term $\Delta V_1$ is   
the one loop correction
to the
effective potential in the MSSM~\cite{Gamberini,RadCorr,Pierce}, 
and $s_a$ is the spin of particle $a$. 
The gauge couplings  are subject to boundary conditions
at the unification scale
$\alpha_2(0)= \alpha_G= \frac{5}{3} \alpha_Y(0)$, while if  the soft parameters are universal one has
$m_i^2(0)=m_0^2+\mu_0^2, ~i=1,2;$ and $m_3^2(0)= - B_0\mu_0$.
The breaking of electroweak symmetry occurs when (a) the determinant  
of the Higgs mass$^2$ matrix turns negative and (b) 
the potential   is  bounded  from below; i.e.
(a) $~~m_1^2 m_2^2 -m_3^4<0$, 
 and $(b)  ~m_1^2+m_2^2 -2|m_3^2|>0$.
Minimization of the potential then yields the following relations
(I) $M_Z^2=2(\mu_1^2-\mu_2^2\tan^2\beta)(\tan^2\beta -1)^{-1}$ and
(II) $\sin 2\beta = 2m_3^2(\mu_1^2 +\mu_2^2)^{-1}$,
where $\mu_i^2=m_i^2+\Sigma_i$,
 where $\Sigma_i$  are the loop 
corrections~\cite{RadCorr,Pierce}.
Here $\tan\beta= v_2/v_1$ is the ratio of the Higgs VEVs.
(I) can be used to fix $\mu$
using the experimental value of $M_Z$, and the
 constraint (II) can be used to eliminate  $B_0$ in favor of $\tan\beta$.  
The supergravity model at low energy can then be parametrized by~\cite{ArnowittNath}
 \be m_0,\;\; m_{1/2},\;\; A_0, \; \;\tan\beta,\;\; {\rm sign} (\mu) \label{msugra}~.\ee 
 After specifying the high scale soft breaking parameters, one implements
  renormalization group analysis (see~\cite{mv} for the two loop analysis) and is then able to predict
  all 32 sparticles masses as well as their couplings and interactions. The full analysis can be done 
  via~\cite{SuSpect}.

 \section{ATLAS and CMS Constraints on Dark Matter Direct Detection in  minimal Supergravity }

  \begin{figure}[h!]
   \begin{center}
       \includegraphics[scale=0.4]{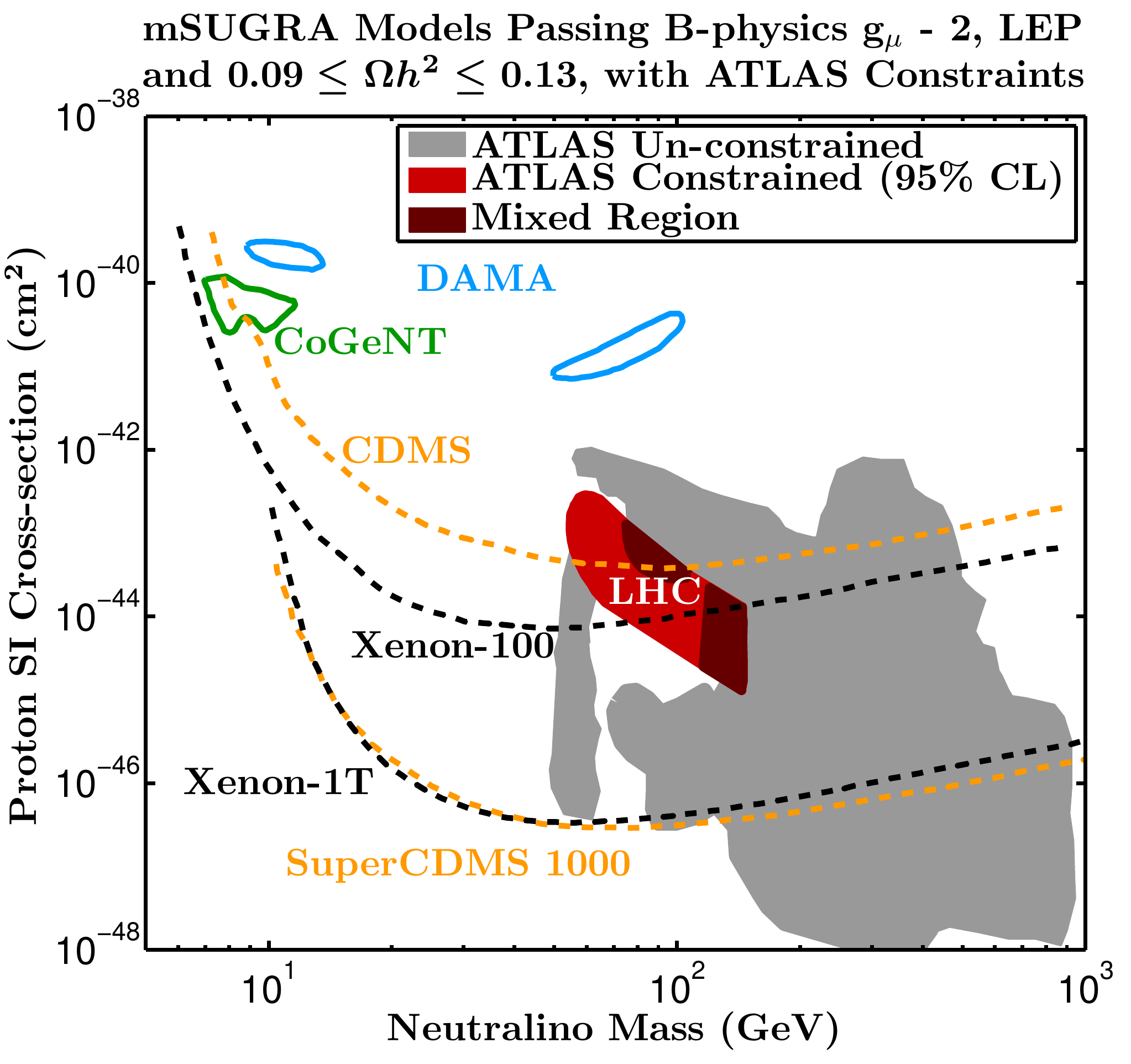}
    \caption{\label{fig1} (color online) 
    A plot of spin independent neutralino-proton cross section
 vs neutralino mass for mSUGRA under experimental constraints. 
The search for supersymmetry at LHC with 35~pb$^{-1}$ luminosity has excluded 
a significant number of models in this signature space which are marked by red color.  
In the red region, all the models in our scans have been constrained by the ATLAS search, 
while in the mixed region (maroon), about 60\% of the models in our scans are constrained by the 
ATLAS search.
We also display the present CDMS~\cite{cdms} and XENON-100~\cite{xenon} curves as well as the future projected experimental curves~\cite{futureXENON,futureSCDMS}.}
\end{center}
\end{figure}

We discuss now the implications of ATLAS and CMS results on dark matter.  For a sample of  works on dark matter and LHC,  we refer the reader to~\cite{darklhc}. 
 SUGRA models predict a dark matter candidate which over much of the parameter
 space is  the lightest neutralino,  the lightest (R-parity odd) superpartner (LSP). 
The LSPs  are traveling with non relativistic speed order $0.001 c$ 
in the galactic halo. This then translates
into the fact that their momentum transfer is very small (order 100 MeV for LSP masses of order 100  GeV)
in collisions with nuclei in a terrestrial detector. 
As such, the relevant interactions for the direct
detection of LSP dark matter  is  calculated in the limit of zero momentum transfer in collisions with nuclei.
For SUGRA models
the interaction Lagrangian is given by~\cite{Chattopadhyay:1998wb,Ellis:2000ds}
\beqn
{\cal L} = \bar{\chi} \gamma^\mu \gamma^5 \chi \bar{q_{i}} 
\gamma_{\mu} (\alpha_{1i} + \alpha_{2i} \gamma^{5}) q_{i} +
\alpha_{3i} \bar{\chi} \chi \bar{q_{i}} q_{i} + \non
\alpha_{4i} \bar{\chi} \gamma^{5} \chi \bar{q_{i}} \gamma^{5} q_{i}+
\alpha_{5i} \bar{\chi} \chi \bar{q_{i}} \gamma^{5} q_{i} +
\alpha_{6i} \bar{\chi} \gamma^{5} \chi \bar{q_{i}} q_{i}~.
\label{lagr}
\eeqn
The spin independent (SI) 
cross section for neutralinos scattering elastically off target nuclei is mostly
 governed by the operator $\alpha_{3i} \bar{\chi} \chi \bar{q_{i}} q_{i}$. 
For heavy nucleus targets, the SI cross section add up coherently
\be
\sigma_{\chi T}  = \frac{4 \mu^2_{\chi T}}{\pi} (Z f_p + (A-Z) f_n)^2~,
\ee
where $\mu_{\chi T}$ is the reduced mass of the neutralino and the target system, and 
$(Z,A)$ are the atomic (number, mass) of the nucleus.
The interactions between the LSP and the target nuclei 
occur  dominantly via 
$t$-channel CP-even Higgs exchange, and $s$-channel squark exchange.
The relevant interactions are given in terms of 
\be
f_{p/n}=\sum_{q=u,d,s} f^{(p/n)}_{T_q} a_q \frac{m_{p/n}}{m_q} + \frac{2}{27} f^{(p/n)}_{TG} \sum_{q=c,b,t} a_q  \frac{m_{p/n}}{m_q}~.
\ee
Here $f^{(p/n)}_{T_u}$, $f^{(p/n)}_{T_d}$, $f^{(p/n)}_{T_s}$ are the nucleon parameters 
which can be obtained from the measurements of the pion-nucleon sigma term, and 
$f^{(p/n)}_{TG}\equiv 1 -f^{(p/n)}_{T_u}-f^{(p/n)}_{T_d}-f^{(p/n)}_{T_s}$. 
Numerical values and further details 
are given in, for example, in Ref.~\cite{micro}. 
The  spin independent
cross section depends sensitively on 
LSP  neutralino  decomposition in terms of its
Bino, Wino and Higgsino eigen components 
($(\tilde{B},\tilde{W}^3) \equiv (\lambda_Y,  \lambda^3)$)
\beqn
\chi \equiv \chi_1^0 &=& n_{11}\tilde{B}  + n_{12} \tilde{W}^3 +n_{13} \tilde{H}_1 +n_{14}\tilde{H}_2~ .
\eeqn
The relevant couplings that enter in the spin independent cross section are~\cite{Chattopadhyay:1998wb,Ellis:2000ds}
\begin{eqnarray}
\label{aq}
a_{q}\equiv a_{3i} & = & - \frac{1}{2(m^{2}_{1i} - m^{2}_{\chi})} \Re \left[\left( X_{i} \right) \left( Y_{i} \right)^{\ast} \right] 
- \frac{1}{2(m^{2}_{2i} - m^{2}_{\chi})} \Re \left[\left( W_{i} \right) \left( V_{i} \right)^{\ast} \right] \nonumber \\
& &  - \frac{g_2 m_{q}}{4 m_{W} B} \left[ \Re \left( 
\delta_{1} [g_2 n_{12} - g_Y n_{11}] \right) D C \left( - \frac{1}{m^{2}_{H}} + 
\frac{1}{m^{2}_{h}} \right) \right. \nonumber \\
& & \ +  \Re \left. \left( \delta_{2} [g_2 n_{12} - g_Y n_{11}] \right) \left( \frac{D^{2}}{m^{2}_{h}}+ \frac{C^{2}}{m^{2}_{H}} \right) \right]~.
\end{eqnarray} 
Here the various quantities $X_i, Y_i, W_i$ etc are defined in~\cite{Chattopadhyay:1998wb,Ellis:2000ds},
where the full forms of $a_q$ can also be found. 
The first two terms arise from squark  $(m_{1i}, m_{2i})$  exchange 
while the remaining terms arise from Higgs exchange which
are almost always dominant in the models we discuss.
The parameters $\delta_{1,2}$ depend on eigen components of the LSP
wave function and $B,C,D$ depend on VEVs of the Higgs fields and the Higgs mixing 
parameter $\alpha$ and are given by
\beqn
{\rm for~u~quarks:}~~\delta_{1} = n_{13}~~~ \delta_{2} = n_{14} ~~~  B = \sin{\beta} ~~~  C = \sin{\alpha} ~~~  D = \cos{\alpha}\\
{\rm for~d~quarks:}~~\delta_{1} = n_{14} ~~~\delta_{2} = -n_{13} ~~~ B = \cos{\beta} ~~~ C = \cos{\alpha} ~~~  D = -\sin{\alpha}~.
\eeqn

  In Fig.(\ref{fig1}) we give the spin independent cross sections vs the neutralino mass 
  after experimental constraints are applied (discussed in Sec.(\ref{cccc})) as well as 
  constraints from the LHC SUSY searches~\cite{Akula:2011zq}.
  We describe the simulations further  in what follows. 
  Also shown  are the XENON-100~\cite{xenon},  CDMS~II~\cite{cdms} and  projected XENON and SuperCDMS
 limits for comparison~\cite{futureXENON,futureSCDMS}. The direct mapping of the parameter space
 constrained by the recent CMS and ATLAS searches is substantial in the spin independent scattering cross section - dark matter mass
 plane. This is achieved  by simulating the LHC SUSY production of the models and SM backgrounds under CMS and ATLAS cuts.  We extend their
 results by considering a larger class of models over the parameter space relevant to early SUSY searches. In Fig.(\ref{fig1}), we identify the  region in this plane that the LHC data constrains.  We will see that this corresponds to the low mass branch of the slepton coannihilation region, 
defined by  $(m_{\tilde l} -m_{\na})/m_{\na}\lesssim 0.2$. Thus, observation of dark matter in the LHC depleted region 
may indicate the presence of nonuniversalities. 
   We discuss now the CMS and ATLAS analyses, and their generalizations and implications in more detail.


\section{LHC  Analysis}
Here, we analyze the nature of the NLSP in the regions of the parameter space depleted 
by the CMS and ATLAS results as well as the SUSY event rates in 
the region that would be
accessible to both the dark matter direct detection experiments and the LHC in the  next rounds of 
data.  As evident from the results of~\cite{atlas0lep,AtlasSUSY, cmsREACH} the 
0 lepton ATLAS analysis is the most stringent, so we mainly focus on this search in our analysis, 
but we have still checked these models with the 1 lepton ATLAS search and the CMS $\alpha_{\rm T}$ 
jet search. We discuss in detail the 0 lepton ATLAS search only; the reader is directed to~\cite{AtlasSUSY, cmsREACH} 
for a more detailed discussion on the other LHC SUSY searches.

We follow the preselection requirements that ATLAS reports in~\cite{atlas0lep, atlasTDR}. Jet 
candidates must  have $p_{T}>20\GeV$ and $\left|\eta\right|<4.9$ and electron candidates must 
have $p_{T}>10\GeV$ and $\left|\eta\right|<2.47$.  Events are vetoed if a ``medium" electron~\cite{atlasTDR} 
is in the electromagnetic calorimeter transition region, $1.37<\left|\eta\right|<1.52$.  Muon candidates must 
have $p_{T}>10\GeV$ and $\left|\eta\right|<2.4$. 
Further, jet candidates are discarded if they are within $\Delta R=\sqrt{(\Delta \eta)^2+(\Delta \phi)^2}=0.2$ 
of an electron.   For the analysis, the (reconstructed) missing energy, $\met$, for an event is the 
negated vector sum of the $p_T$ of all the jet and lepton candidates.  

The analysis is made up of 4 regions, ``A", ``B", ``C" and ``D", each having 0 lepton candidates.  When referring to different cuts in these regions we 
define cuts on the ``selected" jets to mean that the ``selected" jet candidate has $\left|\eta\right|<2.5$ and the bare minimum number of jets in this region 
must satisfy the  requirement. For regions A and B   ``selected" jets refers to the first two hardest jets in the $\left|\eta\right|<2.5$ region and for regions C and D  ``selected"  jets refers to the 
first three hardest jets in the $\left|\eta\right|<2.5$ region. Events are required to have $\met>100\GeV$ and the selected jets 
must each have $p_{T}>40\GeV$ with the hardest jet $p_{T}>120\GeV$. 
Further, 
events are rejected if the missing energy points along the same direction as any of the selected jets., i.e. we require $\Delta \phi \left(j_{i},\met\right)>0.4$, 
where $i$ is over the ``selected" jets. Region A requires events to have $\met>0.3\meff$ with $\meff>500\GeV$ and  regions C 
and D both require events to have $\met>0.25\meff$ with region C requiring $\meff>500\GeV$ 
and region D requiring $\meff>1~{\rm TeV}$. In this case $\meff$ is defined to be the scalar 
sum of the  missing energy and the $p_T$ of the ``selected" jets. 
As in the analysis of~\cite{Akula:2011zq}  we 
do not apply the cut for region B, i.e.  $m_{\rm T 2}>300\GeV$, since the models constrained 
in this region are already constrained in region D~\cite{atlasWeb}. 

For our analysis, we use the simulated SM background of~\cite{Peim} 
  which was generated with {\tt MadGraph 4.4}~\cite{lhcsim} for  parton level processes, {\tt Pythia~6.4}~\cite{lhcsim2} for  hadronization and {\tt PGS-4}~\cite{lhcsim3} for detector simulation. A more thorough discussion on the details of this background can be found in~\cite{Peim,Peim2} and Ref.~1~of~\cite{related},
(see also~\cite{Lessa,Kane:2011zd,cgklr} for discussions on SM background for $2\to N$ processes).  After applying the LHC SUSY analysis to our SM background we are able to reproduce their reported standard model Monte Carlo results.

\section{Result of dark matter analysis with CMS-ATLAS Constraints \label{cccc}} 
We discuss now the implications 
of the  data from CMS and ATLAS on dark matter. To this end we first carry out a survey of the
mSUGRA parameter space as follows: 
$m_0 \in (10, 4000)$ GeV, 
$m_{1/2} \in (10,2000)$ GeV, 
$A_0 \in (-10,10)m_0$, 
$\tan\beta \in (1,60)$.  
Performing a 
 general survey of the mSUGRA model space we simulate the models that satisfy radiative electroweak symmetry breaking 
 (REWSB) as well as direct and indirect experimental constraints  including
  sparticle mass limits, B-physics constraints,  and  constraints from $g_{\mu} -2$.
  We further require that the relic density be within the observed WMAP limit~\cite{WMAP}, $0.0896 < \Omega_\chi h^2 < 0.1344$.
     These indirect constraints were calculated using {\tt MicrOmegas}~\cite{micro}, with the Standard Model contribution in the ${\mathcal{B}r}\left(b\to s\gamma\right)$ corrected using the NNLO analysis of 
 Misiak~{\it et al}.~\cite{Misiak:2006zs,Chen:2009cw}.
We apply the following ``collider/flavor constraints"~\cite{pdgrev} 
  $m_h > 93.5~\GeV$, $m_{\sta} > 81.9~\GeV$, $m_{\cha} > 103.5~\rm GeV$, $m_{\ta} > 100~\GeV$, $m_{\ba} > 89~\GeV$, $m_{\ser},m_{\sel} > 107~\GeV$, $m_{\smr},m_{\sml} > 94~\GeV$, and $m_{\g} > 400~\GeV$, along with
   $\left(-11.4\times 10^{-10}\right)\leq
\delta \left(g_{\mu}-2\right) \leq \left(9.4\times10^{-9}\right)$, see~\cite{Djouadi:2006be},
${\mathcal{B}r}\left(B_{s}\to \mu^{+}\mu^-\right)\leq 4.2\times10^{-8}$ (90\%~C.L.)~\cite{Abazov:2010fs}, and 
$\left(2.77\times 10^{-4} \right)\leq {\mathcal{B}r}\left(b\to s\gamma\right) \leq \left( 4.37\times 10^{-4}\right)$~\cite{bphys}. 
 
  \begin{figure*}[t!]
   \begin{center}
      \includegraphics[scale=0.31]{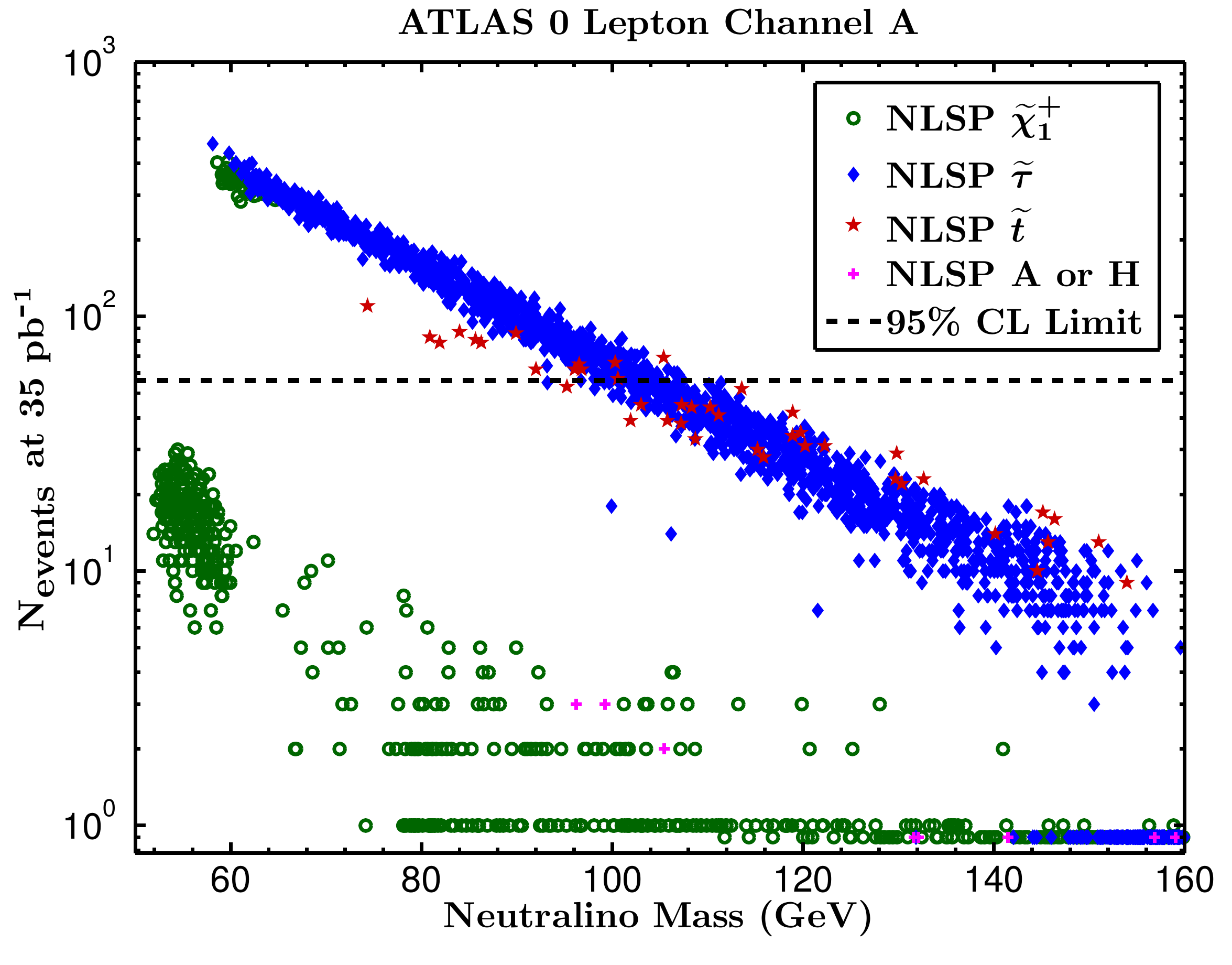}
   \includegraphics[scale= 0.31]{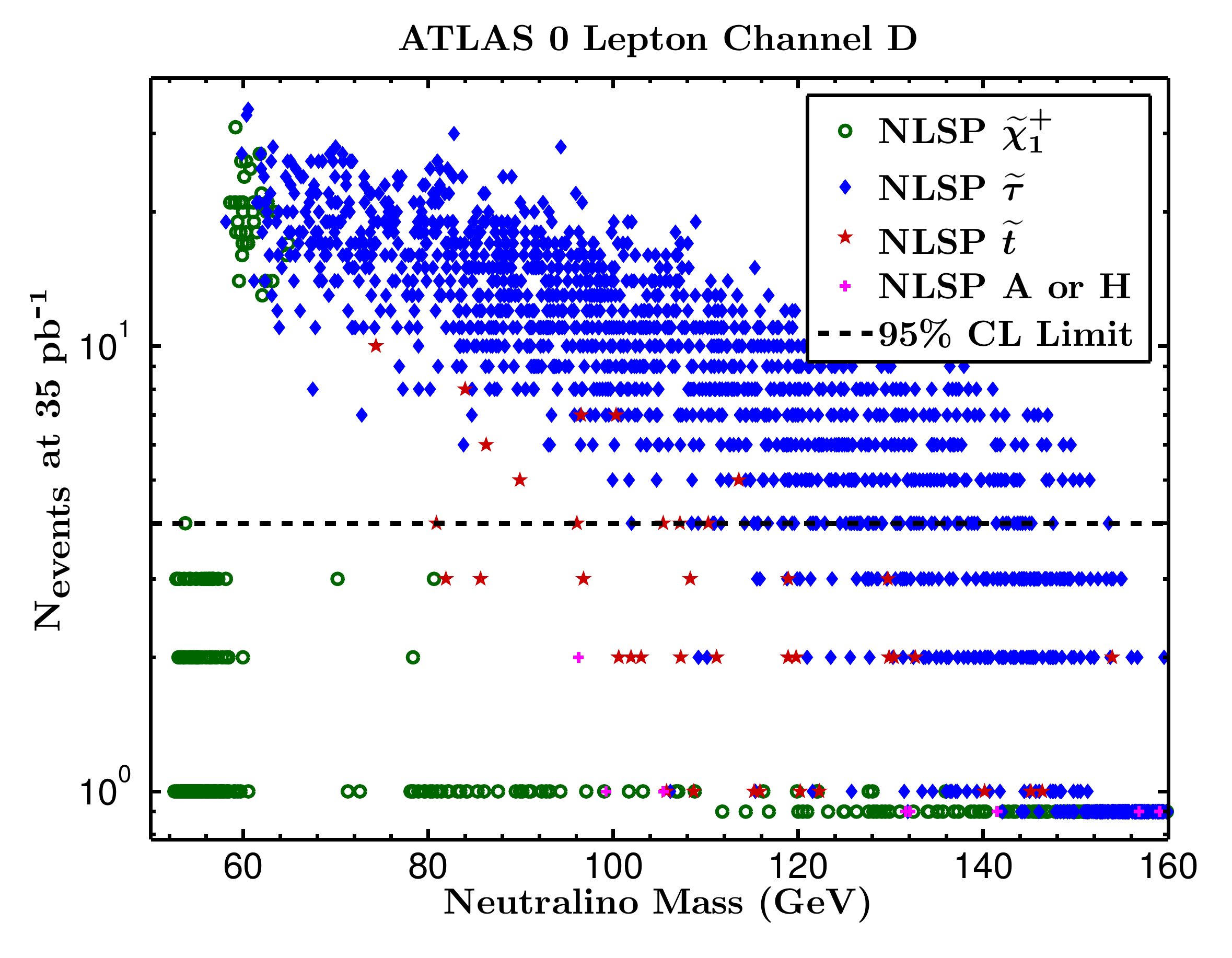}
\caption{\label{fig2} (color online) 
Exhibition of the number of SUSY events in the ATLAS 0 lepton analysis  and the corresponding NLSPs against the neutralino mass with 35~pb$^{-1}$ of integrated luminosity
for a subset of models around the LHC excluded region of Fig.(\ref{fig1}). 
Left panel: Region A~\cite{atlas0lep};  Right panel: Region D~\cite{atlas0lep}.
The dashed black lines can be viewed as the 95\%~C.L. limit in each signal region, as they correspond to the event thresholds reported by ATLAS along the the $m_{0}-m_{1/2}$ boundaries~\cite{atlasWeb}.
Essentially, the models being eliminated by the ATLAS results (above the dashed black line) are those with the stau as the NLSP. 
}
\end{center}
\end{figure*}

\begin{figure}[t!]
   \begin{center}
        \includegraphics[scale=0.35]{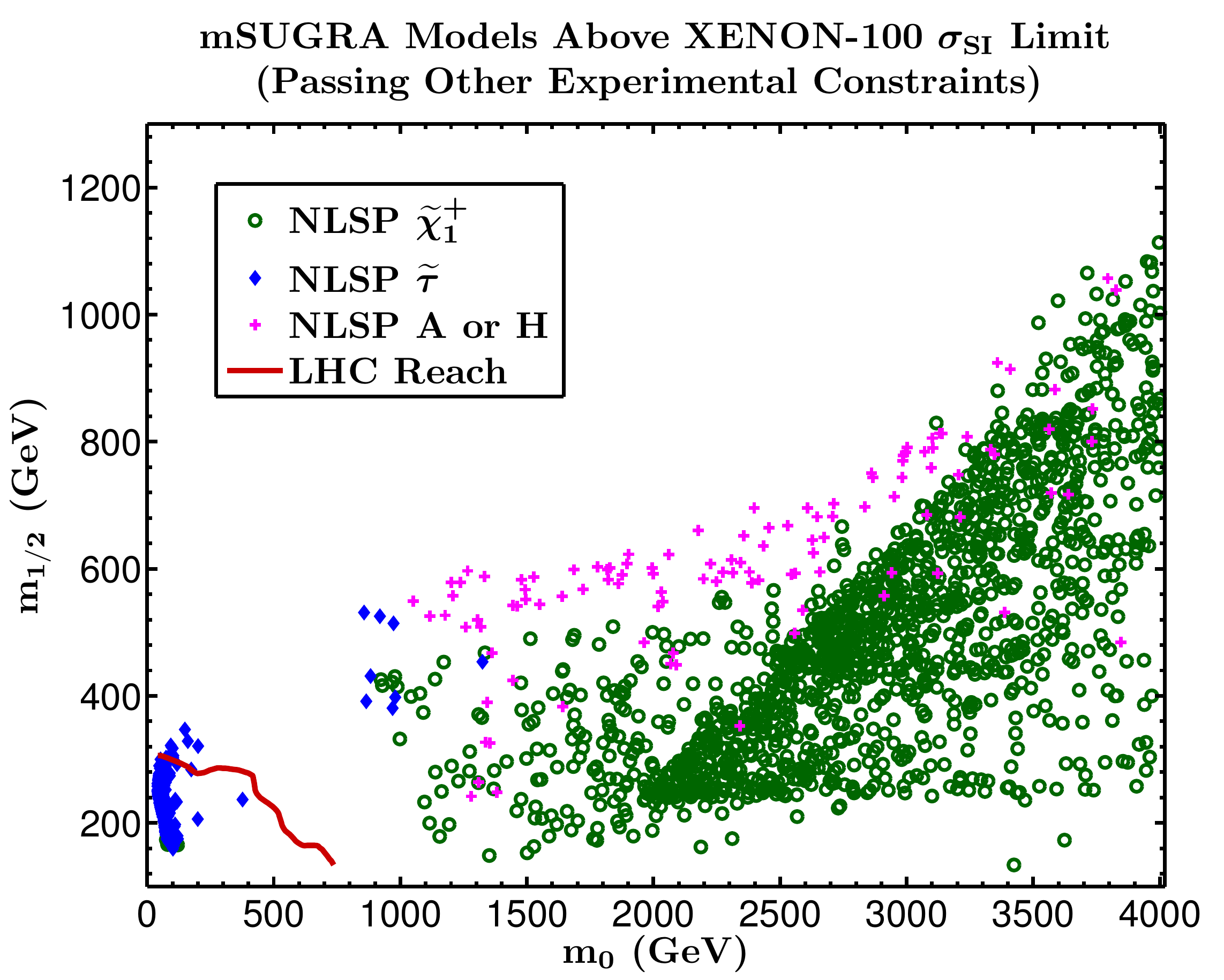}     
        \includegraphics[scale=0.35]{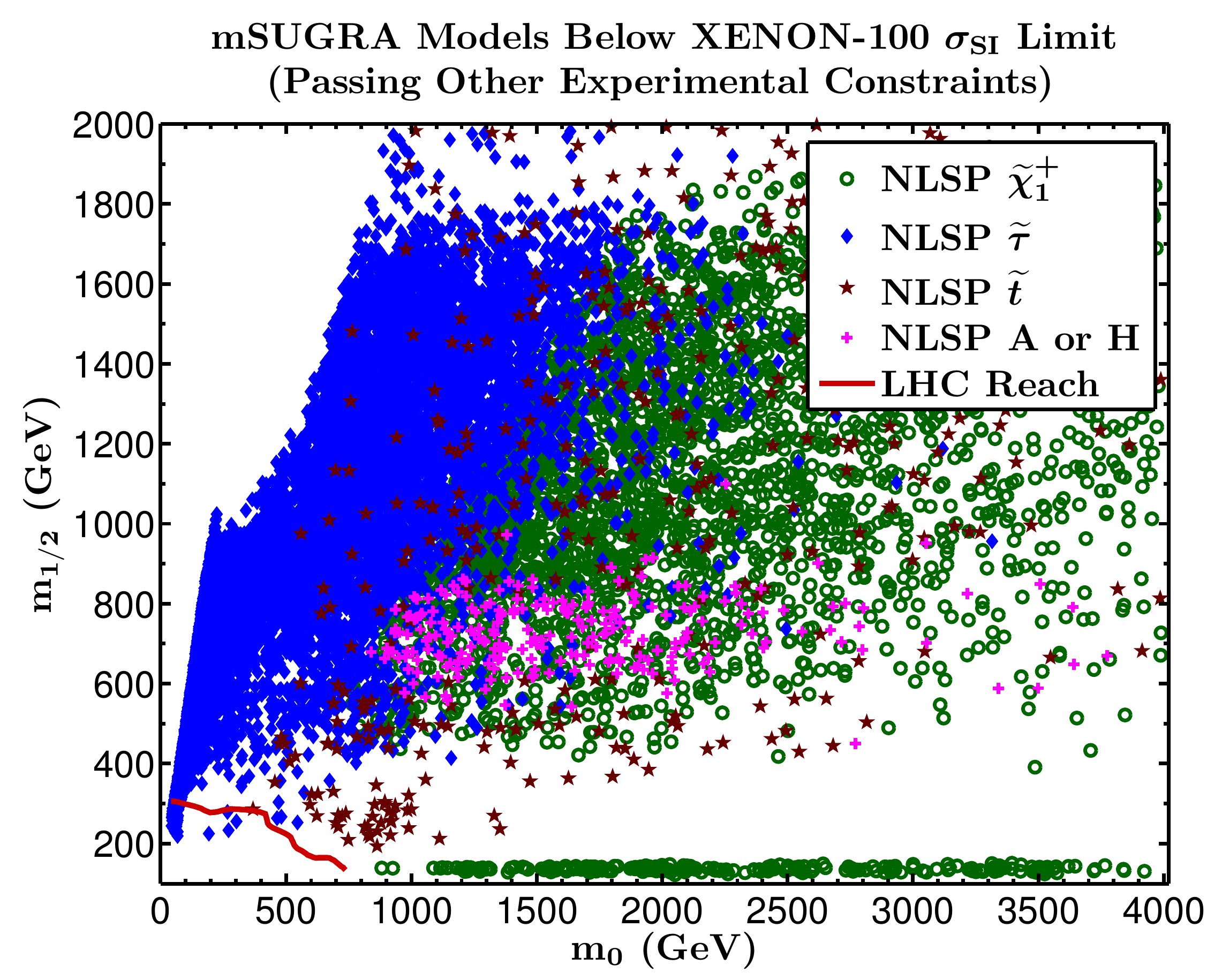}     
             \caption{\label{figplace} (color online) 
Exhibition of models in the $m_{0}-m_{1/2}$ plane denoted by their NLSPs and the ATLAS 0 lepton curve (red) is drawn for comparison (see Fig.(\ref{fig1})).  The left panel corresponds to the models that have been constrained by XENON-100~\cite{xenon} and the right panel corresponds to the models that are unconstrained by XENON-100. All models have the same constraints as Fig.(\ref{fig1}).
From this analysis we see explicitly that  the reported XENON constraints are severe in the larger $m_0$ region constraining  the hyperbolic branch, 
while the low $m_0$ region, which are  the low mass
slepton coannihilation regions, are  being constrained by both XENON and the LHC. 
}
\end{center}
\end{figure}

To investigate the constraints from the LHC SUSY search on the dark matter detection signals, 
we scanned over 20 million models in the mSUGRA parameter space. After imposing the various 
experimental constraints as previously discussed, we simulate the models with the ATLAS 0-lepton 
analysis. It is found that there exists a large portion of the signature space in the 
spin independent cross section-neutralino plane which is being excluded by the ATLAS 
0-lepton search. This excluded region which is marked by red color as shown in Fig.(\ref{fig1}) 
was populated by mSUGRA models before considering the new LHC data. We further 
divide the excluded region into the red region where all the mSUGRA models scanned 
are excluded by the LHC data, and the two maroon regions each with about 60\% of the models 
excluded by the LHC. 
(Note that ATLAS carried out their analysis for a few fixed values of $\tan\beta$ and $A_0$ while our analysis allow these to vary.) 
Next, by considering the NLSP, we find that essentially all of the region that is depleted by the LHC 
at 95\% CL is the low mass region of the slepton coannihilation branch.

 This is shown more clearly in 
  Fig.(\ref{fig2}) where we display the number of SUSY events vs the neutralino mass for a subset of models in the two panels
 corresponding to the regions A and D with low neutralino masses. We do not display region C since it gives results similar 
 to region A and we do not display region B since it is subsumed in region D.  
The dashed black lines in Fig.(\ref{fig2}) can be viewed as the 95\%~C.L. limit in each signal region, 
as they correspond to the event thresholds reported by ATLAS.
Indeed, most of  the model points being 
 constrained by the LHC are those where the stau is the NLSP appropriate for the slepton coannihilation 
 branch. Further, very few of the model points  are constrained by the ATLAS analysis 
  which lie on the  Hyperbolic Branch (HB) (Focus Point region)~\cite{HB} of REWSB. The NLSP on the HB
 is mostly the light chargino and from Fig.(\ref{fig1}) we find that  very few 
 of the chargino NLSP models are currently constrained by the ATLAS analysis.

In contrast,  the higher mass HB/FP region is becoming constrained by the XENON data~\cite{xenon}.
This effect can be seen in Fig.(\ref{figplace}) where we show the $m_{0}-m_{1/2}$ plane for the mSUGRA
case denoted by their NLSP where the models on the left panel are constrained by XENON-100 and the models on 
the right panel are unconstrained by XENON-100.  

Thus, we come to the conclusion that the ATLAS constraints are very severe for the low
$m_0$ region, while the XENON constraints are very severe for the large $m_0$  region as shown.  
As can be seen from Ref.~\cite{Akula:2011zq}, the region which is now being constrained by XENON corresponds to $\mu \lesssim 400 ~\rm GeV$
and here the LSP wavefunction has a significant Higgsino component.
We add here that  bulk region and the higgs pole region (the latter being the horizontal strip of essentially fixed  $m_{1/2} \sim O(100-150) ~\rm GeV$) remain largely untouched by either experiments.

More generally while the recent XENON analysis~\cite{xenon} has presented plots
along with  mSUGRA~\cite{ArnowittNath} (see Eq.~(\ref{msugra})) model points on top of the data --
we  suggest that the XENON collaboration include the  50~GeV to 65~GeV  mass range  of mSUGRA
 in their
constraint plots  as this is the region where the XENON
data shows its  greatest  present sensitivity (see e.g. Ref.~1~of~\cite{related} for this dense
region of parameter space; the Higgs pole region mentioned above).
We also remark that in the analysis of the spin independet cross section we  used the  default values of the form factors as given in Ref.~\cite{micro}.
It is well known, the predictions for the SI cross section
are sensitive to the precise knowledge of the form factors and in particular  the strange quark form factor. In addition, variations on the order of 5 or larger have been reported  in the second reference  of~\cite{Ellis:2000ds} and in~\cite{micro}  over a reasonable range of the pion-nucleon sigma term (for which the above form factors depend on).
These uncertanties should be kept in mind while interpreting the results of dark matter direct detection experiments on the parameter space of models.  Thus, while
we have shown
in Fig.(\ref{figplace}) the regions which lie below and above the reported XENON limits one does need to factor in more generally  the uncertainties in the hadronic matrix elements as well as the uncertainties in astrophyscial quantaties to have a more precise account of the constrained region of parameter space. However, such an analysis  goes beyond the scope of this work. Thus our aim here is to emphasize that the sensitivity of the XENON detector is encroaching on a new part of the space of SUGRA models, and it is beginning to provide more stringent constraints
on the larger $m_0$ region for  which the Higgsino component of the LSP wavefunction can become significant.

\section{SUGRA models with non-universal  breaking} 
The analysis    for the mSUGRA case highlighted in Fig.(\ref{fig1}) 
shows a deficit of models after the LHC constraints are  applied in the region  under the XENON-100 curve 
   in the neutralino
mass range of $50\GeV$ to $100\GeV$ corresponding to the  slepton coannihilation region.
While the assumption of 
universal boundary conditions on soft breaking in supergravity grand unification~\cite{sugra} 
is the simplest possibility leading to the model mSUGRA, the framework of supergravity unification~\cite{sugra}
allows for non-universalities in the soft parameters which occurs generically for several
classes of string motivated models (see~\cite{Ibanez,Corsetti, string2,Nath:1997qm}). 

\begin{figure}[t!]
   \begin{center}
        \includegraphics[scale=0.4]{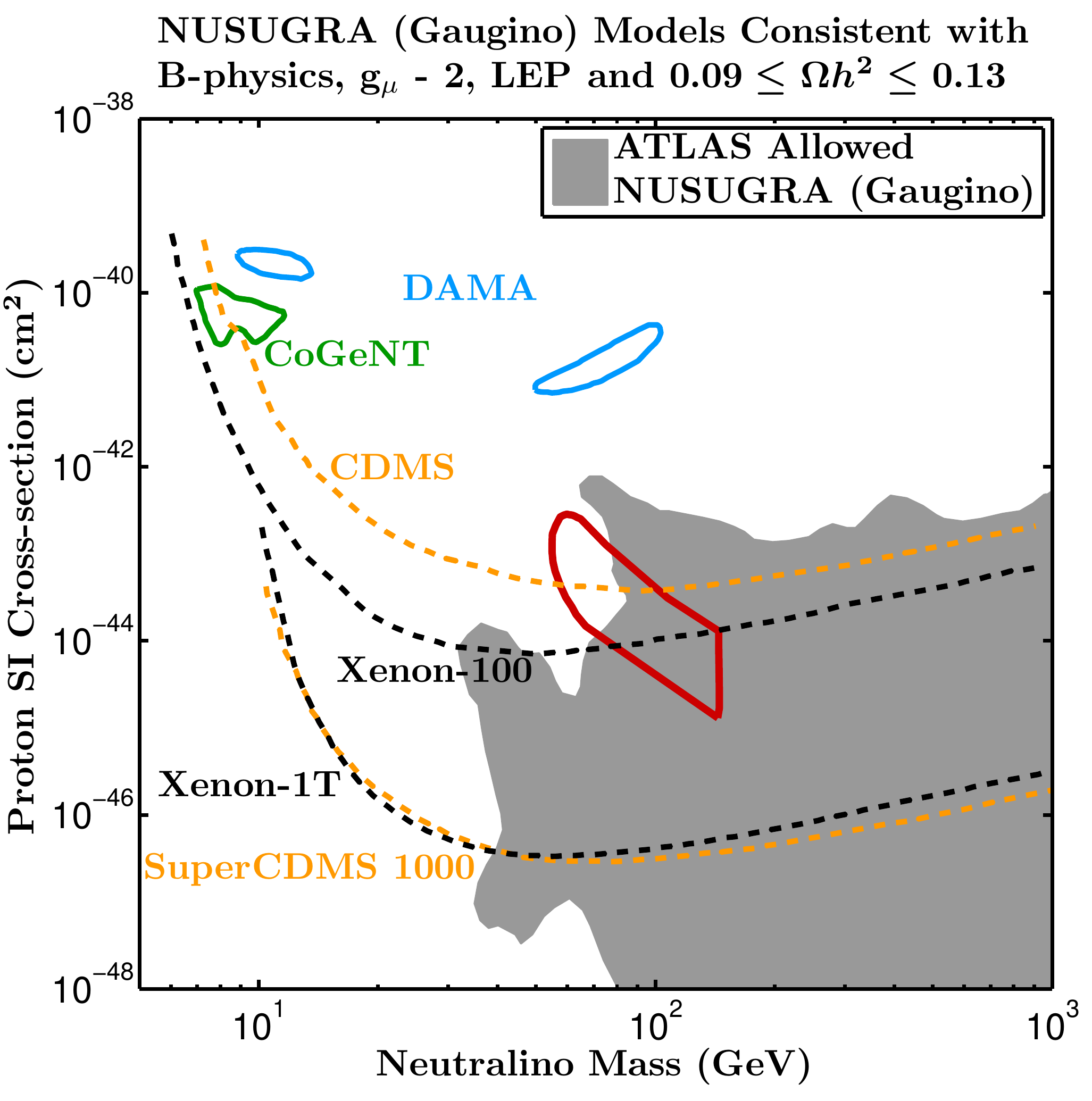}       
             \caption{\label{fig3} (color online) 
Analysis of models with non-universalities in the gaugino masses with the LEP, Tevatron,
$g_{\mu}-2$, FCNC and WMAP constraints. The red contour 
is the region depleted for mSUGRA by the ATLAS results and is shown for comparison.
The random scan does not emphasize the mSUGRA parameter region.
 }
\end{center}
\end{figure}

   \begin{figure*}[t!]
   \begin{center}
  \includegraphics[scale=0.35]{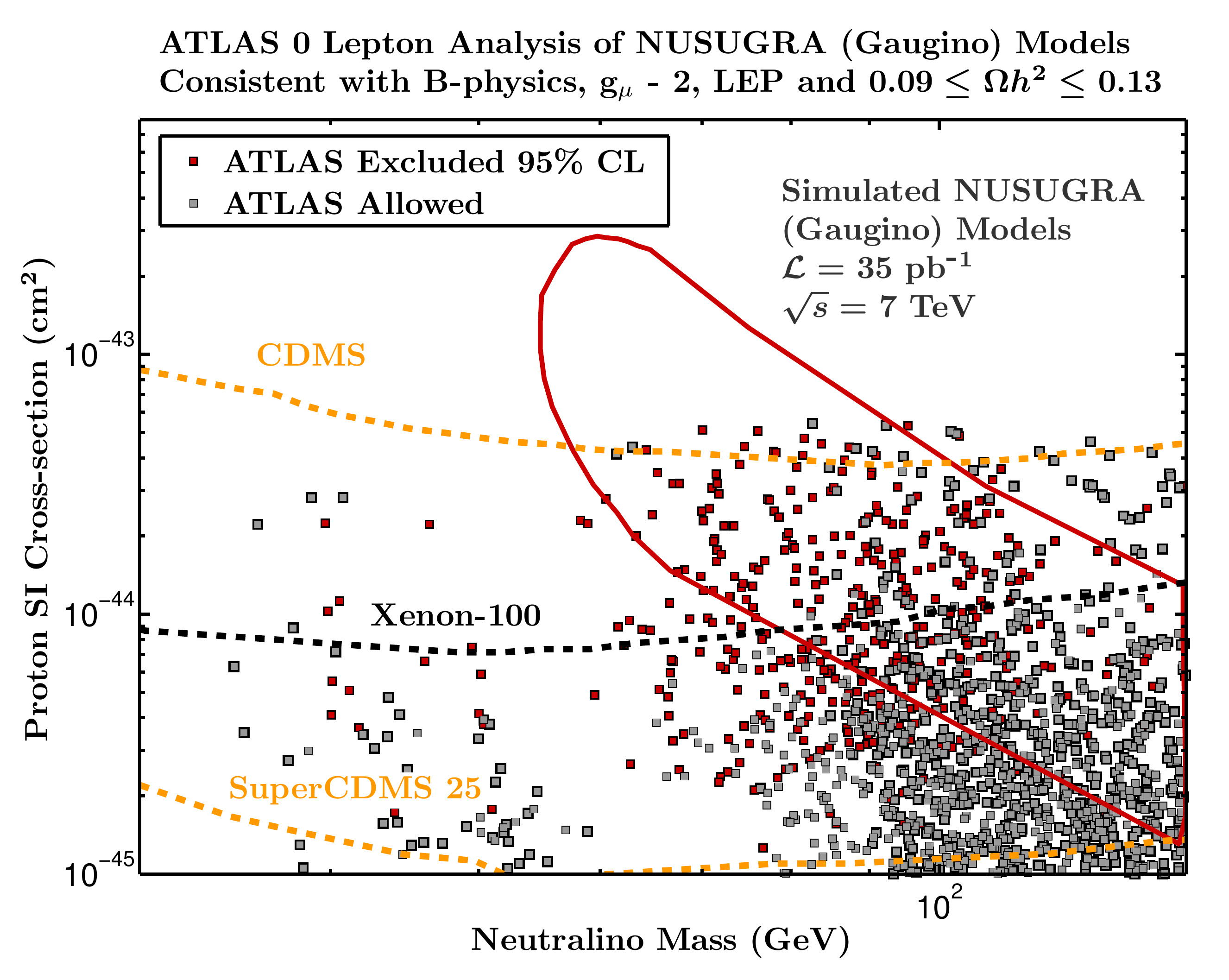}    
                      \includegraphics[scale=0.4]{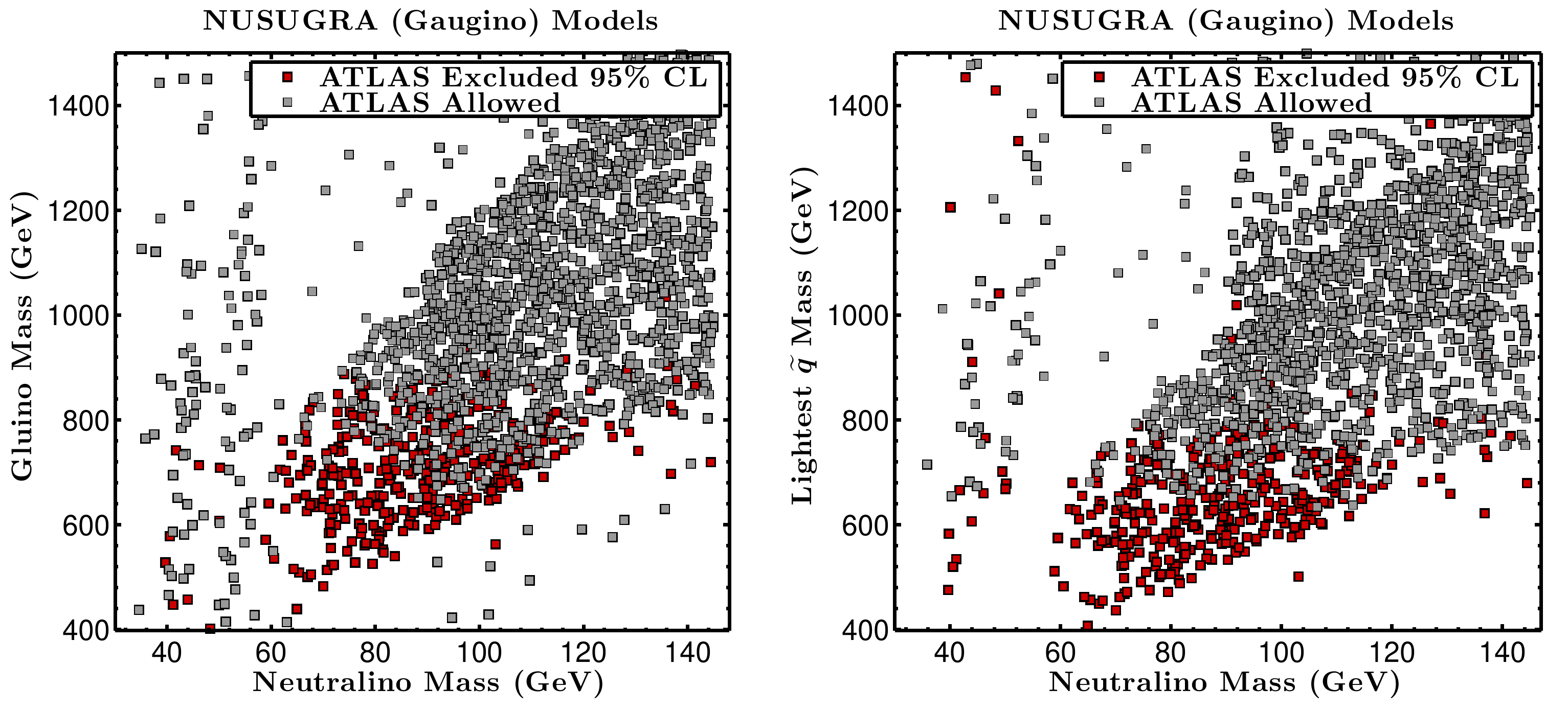}    
             \caption{\label{fig4} (color online) 
Repopulation of the region depleted by ATLAS. Shown are NUSUGRA
models, where the red contour is the ATLAS constrained 
 region in mSUGRA. The non-universal gaugino models simulated (a subset of
 models in Fig.(\ref{fig3}))  under the ATLAS 0 lepton cuts
that are  constrained  by the  analysis indicated by  red squares. 
 The bottom  two panels show the gluino mass  and the lightest second generation squark mass  where we note 
a gluino
 mass as low as 400~GeV and squark masses as low as 600~GeV are unconstrained by the present ATLAS data.
}
\end{center}
\end{figure*}

Non-universal gaugino masses can arise in two ways (a) from tree level supergravity with a gauge kinetic
function dependent on singlets or products of singlets and fields which transform 
under the gauge groups of the standard model  (b) from loop induced gaugino masses
dependent on the beta function coefficient for each group.  
For tree level gaugino masses one has 
\beqn
M_a = \frac{1}{2\Re(f_a)} F^I \partial_I f_a \ .\nonumber
\eeqn
where $F^I$ are the order parameters of SUSY breaking, $I$ denotes
the hidden sector (singlet) fields responsible for the breaking of SUSY
and $f_a$ is a diagonal gauge kinetic function, where $a$ is an adjoint 
index for each gauge group.
In addition  for loop induced gaugino masses one has~\cite{arnowitt,Ibanez,Gaillard}
\begin{equation}
M^1_{a}|_{\rm adj} = -b^0_a g^2_a m_{3/2} + \dots
\nonumber
\label{mary}
\end{equation}
  where the higher order terms are given in~\cite{Gaillard}
  and the beta function coefficient is given in terms of $C_a, C^i_a$ ;
  the quadratic Casimir operators for the gauge group $G_a$ respectively in the adjoint
representation
\begin{equation}
b^0_a = \frac{1}{16 \pi^2} (3 C_a - \sum_i C^i_a) \nonumber
\end{equation}
Thus 
 we now consider the case of non-universal supergravity~(NUSUGRA) models to see if the  region 
 depleted in the mSUGRA case 
 can become populated when non-universalities are included. 
 Here we will keep the analysis rather general and parametrize the non-universalites as 
in the gaugino masses which can be sourced from tree level supergravity, from loop induced gaugino masses,
and most generally a combination of both as
\be M_a = m_{1/2} \ (1+\delta_a) \ee
 at the GUT scale for the gauge groups $U(1), SU(2)_L, SU(3)_C$ 
corresponding to $a=1,2,3$.  The ranges chosen  
are $\delta_a=(-1,1)$ with the ranges for the remaining parameters as in the mSUGRA case.

The result of the analysis is shown in Fig.(\ref{fig3}) where we exhibit the allowed set of  models  
over a broad range of neutralino masses which satisfy all the experimental constraints, but do not yet have the LHC SUSY search constraints applied to them. 
The area  depleted by the LHC for the mSUGRA case lies within the red boundary and is shown
for comparison.
One observes that the presence of non-universalites in the gaugino sector repopulates
a significant part of the region
of the signature space 
in the spin independent scattering cross section-neutralino mass plane that is 
constrained by the LHC SUSY searches relative to the case of minimal
SUGRA. This region of repopulation is found to produce a consistent relic
density via multiple coannihilation channels. 

 In particular, because the chargino mass can be split from the LSP
mass with non-universalites in the gaugino sector consistent with the LEP bound
on the chargino mass,  the low mass region below
the light CP even Higgs pole, which is largely the  Z-pole region, is now allowed by the
relic density constraint. Thus one can have a dark matter mass as low as 
\be  m_{\na} \gtrsim 40 ~ \rm GeV ~~~(NUSUGRA-gauginos) \ee
in the NUSUGRA case, where the lower limit is higher in the mSUGRA case to be consistent with the LEP data.

The top panel of Fig.(\ref{fig4}) gives the analysis 
 with a focus on the  $50\GeV$ to $100\GeV$ neutralino mass region where we also apply
the LHC analysis as already described. From Fig.(\ref{fig3}) and the top panel of Fig.(\ref{fig4}), it is apparent 
that  the gaugino mass non-universalities produce a significant repopulation of the region with models specifically
 in 
 the $50\GeV$ to $100\GeV$ neutralino mass range.
 Also shown in the bottom two  panels of Fig.(\ref{fig4}) are 
    the gluino mass and the 
     lightest second generation squark mass. We note that a gluino
 mass as low as 400 GeV and a squark mass as low as 600 GeV are unconstrained by the present ATLAS data.
Similar results are obtained when non-universalities in both the 
gaugino sector and the Higgs sector~\cite{Nath:1997qm} are present. In this case the analysis 
gives results similar to those of Fig.(\ref{fig4}) with
 a larger density of allowed models which populate the region depleted by the LHC SUSY searches.


\section{Conclusion} 
The implications of the first SUSY analysis by CMS and ATLAS 
on 
 supersymmetric dark matter are analyzed. It is found that the CMS and ATLAS  constraints
deplete a significant branch of the slepton coannihilation regions in the mSUGRA parameter space
where dark matter can originate in the early universe while the Higgs pole region and the Hyperbolic Branch (focus point region) 
are not constrained.  However, a large portion of the  Hyperbolic Branch region is now becoming constrained by the recent XENON data. The effect of non-universalities in the gaugino masses  are analyzed and it 
 is found that a part of the region in the 
spin-independent cross section vs the LSP mass
 plane depleted by the 
 CMS and ATLAS analysis for mSUGRA is repopulated when non-universalities are included, i.e.,
 for the  NUSUGRA case.
Thus observation of dark matter in the mSUGRA  region depleted by the ATLAS constraints 
could point to supergravity models with  non-universal  soft breaking.


\section{Acknowledgements}
\noindent  We would like to thank Darien Wood for discussions 
 regarding the likelihood-based approach for setting limits  on new physics.
 This research is  
supported in part by grants  DE-FG02-95ER40899,   PHY-0757959,  PHY-0969739,  
and by  TeraGrid  grant TG-PHY100036.\\
\noindent

{\bf Note added}:
Near the completion of this work  a new ATLAS analysis~\cite{Collaboration:2011ks}
appeared and our results are consistent with their analysis. Further, 
after the appearance  of the work presented here,  an analysis in similar spirit appeared in Ref.~\cite{Farina:2011bh},
and  their  overlapping results are consistent with ours.  For the case of minimal supergravity  Ref.~\cite{Farina:2011bh}
exhibits the NLSPs in the 
spin independent cross section - LSP mass plane.  This is a useful technique
for understanding the physical content of models in this signature space as discussed in~\cite{ZLDFPN,DF}.


\end{document}